\documentclass[]{interact}

\usepackage{epstopdf}
\usepackage[caption=false]{subfig}

\usepackage[numbers,sort&compress]{natbib}
\bibpunct[, ]{[}{]}{,}{n}{,}{,}
\renewcommand\bibfont{\fontsize{10}{12}\selectfont}

\theoremstyle{plain}
\newtheorem{theorem}{Theorem}[section]

\theoremstyle{definition}
\newtheorem{definition}[theorem]{Definition}

\theoremstyle{remark}

\usepackage{multirow}

\begin{document}

\articletype{Original Article}

\title{Multi-state models for double transitions associated with parasitism in biological control}

\author{
\name{Idemauro Antonio Rodrigues de Lara\textsuperscript{a}\thanks{Corresponding author: idemauro@usp.br}, Gabriel Rodrigues Palma\textsuperscript{b},  Victor José Bon\textsuperscript{c}, Carolina Reigada\textsuperscript{c} and Rafael de Andrade Moral\textsuperscript{b}}
\affil{\textsuperscript{a}University of  S\~{a}o Paulo,  Brazil~(``Luiz de Queiroz'' Agriculture College); \textsuperscript{b}Maynooth University, Ireland~(Department of Mathematics and Statistics);
\textsuperscript{c}Federal University of S\~{a}o Carlos,  Brazil~(Department of Ecology and Evolutionary Biology).
}
}

\maketitle

\begin{abstract}
Competition between parasitoids can reduce the success of pest control in biological programs using two species as bio-control agents or when multiple species exploit the same host crop. Parasitoid foraging behavior and the ability to identify already parasitized hosts affect the efficacy of parasitoid species as bio-agents to regulate pest insects. We evaluated the behavioural changes of parasitoids according to the quality of hosts ({\it i.e.}, previously parasitised or not), and the characterisation of these transitions over time via multi-state models. We evaluated the effects of previous parasitism of the brown stinkbug {\it Euschistus heros} eggs on the parasitism rate of the species {\it Trissolcus basalis} and {\it Telenomus podisi}. We successively modelled the choice of eggs (with three possibilities: non parasitised eggs, eggs previously parasitised by {\it T. podisi}, and eggs previously parasitised by {\it T. basalis}) and the conditional behaviour given the choice (walking, drumming, ovipositing or marking the chosen egg). We consider multi-state models in two successive stages to calculate double transition probabilities, and the statistical methodology is based on the maximum likelihood procedure. Using the Cox model and assuming a stationary process, we verified that the treatment effect was significant for the choice, indicating that the two parasitoid species have different choice patterns. For the second stage, i.e. behaviour given the choice, the results also showed the influence of the species on the conditional behaviour, especially for previously parasitised eggs. Specifically, {\it T.podisi} avoids intraspecific competition and makes decisions faster than {\it T. basalis}. In this work, we emphasise the methodological contribution with multi-state models, especially in the context of double transitions.
\end{abstract}

\begin{keywords}
stochastic processes; entomological data; foraging behaviour; stationarity; likelihood procedure, transition intensities.


\end{keywords}

\section{Introduction}

Entomology has a important role in agricultural sciences, specifically because it includes studying and understanding the insect-insect and plant-insect interactions, which can be used to improve agricultural production \cite{gallo2002, Shimbori2023}. Among the many types of studies conducted by entomologists, here we focus on studies related to biological control. This involves utilising living organisms to reduce the population of a target pest species. One example is the host-parasitoid system, where a parasitoid species is used to control the population of a pest that is used as a host (e.g. \textit{Dichelops melacanthus}, \textit{Euschistus heros}, and \textit{Podisus nigrispinus} \cite{queiroz2018}).

Several insects that play an important role in the parasitism of insect pests have been reported in the literature, including the parasitoids \textit{Tamarixia radiata}, \textit{Telenomus podisi}, \textit{Trissolcus basalis}, and several species of the genus \textit{Trichogramma} \cite{camarozano2021}. By studying their controlling capabilities in laboratory and field conditions, we may enhance the efficacy of biological control and consequently reduce the economic damage caused by insect pests. A direct contribution of experiments related to the biology of parasitoids is the estimation of parasitism rates. Several factors can affect them, including hyperparasitism, where a parasitoid of a different species parasitises an egg that has already been parasitised, constituting a competition interaction \cite{yao2022}. However, hyperparasitism can also happen unintentionally when the parasitoid does not detect the eggs already parasitised by its species.

It is common for an entomological study to be longitudinal over time, both in field and laboratory-based studies. Moreover, recorded responses are often categorical. In such cases, the responses may represent behaviours or choices the insects make in different scenarios. An example was presented by \cite{Lara2020} to understand the movement patterns of female adults of {\it Diaphorina citri}, a pest of citrus plantations, having as the response variable the preference for different potted plant positions. 

The parasitoid reproductive behaviour and/or host quality discrimination by competitive parasitoid species, conditional to the host being previously parasitised or not, can be evaluated by recording insect behaviour data (i.e., drumming, ovipositing and marking host eggs) over time. In such cases, the responses may represent behaviours or choices the insects make in different scenarios of host quality. 

It is known that the analysis of longitudinal categorical data can be done using Generalized Linear Models~(\cite{Agresti2010}, \cite{Tutz2012}), such as marginal, mixed effects and transition models~\cite{Diggle2002}. Each of these models has its particularities that depend specifically on the design and objectives of the study. In Entomology, for example, the interaction between species can be measured from changes in behaviour over time under certain experimental conditions. Marginal and mixed effects models cannot model these changes over time. In contrast, transition models are very useful to describe the occurrences from one state to the next and also to assess the effect of experimental design conditions~(\cite{Zeger1992}, \cite{Lara2017}).

Transition models are based on stochastic processes, and a classical reference is \cite{Ware1988}, which distinguishes classes of discrete-time and continuous-time models. When the process is in continuous time, they are also known as state-space models (\cite{meira2009}, \cite{Paula2023}). While in the discrete case, we limit ourselves to evaluating the transition probabilities assuming equally spaced time occasions. For the continuous case, there are options for inferences with respect to time, defined by infinitesimal parameters or intensity rates. Thus, not only are the probabilities of state changes described, but also the intensity with which these changes occur can be modelled, which is more informative.

The focus of this work is on a continuous time transition model~(space-state mode) motivated by a biological problem arising from an experiment involving the parasitoids {\it Telenomus posidi}  and {\it Trissolcus basalis}, which are useful for pest control in soybean~\cite{Bon2021}.  

The egg parasitoids \textit{Telenomus podisi} and \textit{Trissolcus basalis} (Hymenoptera: Scelionidae) are important natural enemies used in biological control programs for different species of soybean pest bugs \cite{de2012}. These parasitoid species are termed generalists because they attack different host species, including the eggs of the brown stinkbug \textit{Euchistus heros} (Fabricius, 1798) (Hemiptera: Pentanomidae) \cite{Correia1991}.

The use of \textit{T. podisi} and \textit{T. basalis} as biological control agents for soybean bugs has occurred through mass releases of these parasitoids in Brazil, and in some regions, both species can also be found in crops \cite{bs1994}. In both situations, competitive interactions between species can be frequent during foraging by hosts, bringing consequences for pest control and the maintenance of these natural enemies in the soybean agroecosystem after parasitoid releases \cite{cusumano2016}.

Competition for hosts can be reduced when parasitoid females are able to discriminate between already parasitised and non-parasitised hosts. This ability to discriminate host conditions is more frequent at the intraspecific level \cite{godfray1994}. The identification of previously parasitised hosts is carried out using chemical and physical clues (semiochemicals and infochemicals) \cite{godfray1994}, to avoid the occurrence of multiparasitism or superparasitism, which directly affects the quality and quantity of the offspring's resources, which can lead to the death of the developing parasitoid larvae \cite{harvey2013}.

Knowledge of the effects arising from competition between parasitoids must be taken into account when defining strategies for the use of these agents in biological pest control programs \cite{follett2000}. In this way, understanding and knowing the biological and behavioural aspects, and also the biodiversity and distribution of parasitoid species in a given location becomes relevant to developing successful biological control programs \cite{correa1993}. In this context, it is also necessary to use appropriate statistical methods, which allow the study of species behaviours over time and at the same time evaluate intraspecific competition. In our motivational study, the parasitoids could make two successive choices, the first before choosing an egg type (non-parasitised, parasitised by their own species or by the opposing species), and then after choosing an egg (marking, ovipositing or drumming on the egg); therefore, the insects presented double transitions over time. In this context, the main goal is to present an extension to the multi-state models associated with successive transitions of the parasitoids as a methodological contribution to understanding the pattern of preferences and behaviours of these species. 

The remainder of this article is structured as follows: in Section \ref{material}, we present our motivational case study; fundamentals of stochastic processes and multi-state models are presented in Section \ref{methodos}; results are presented and discussed in Section \ref{results}; a biological discussion of the results is presented in the Section \ref{discussion}; and finally our final considerations are made in Section \ref{conclusion}.

\section{Case study}\label{material}

\begin{figure}[!ht]\centering
	\includegraphics[width=1\textwidth]{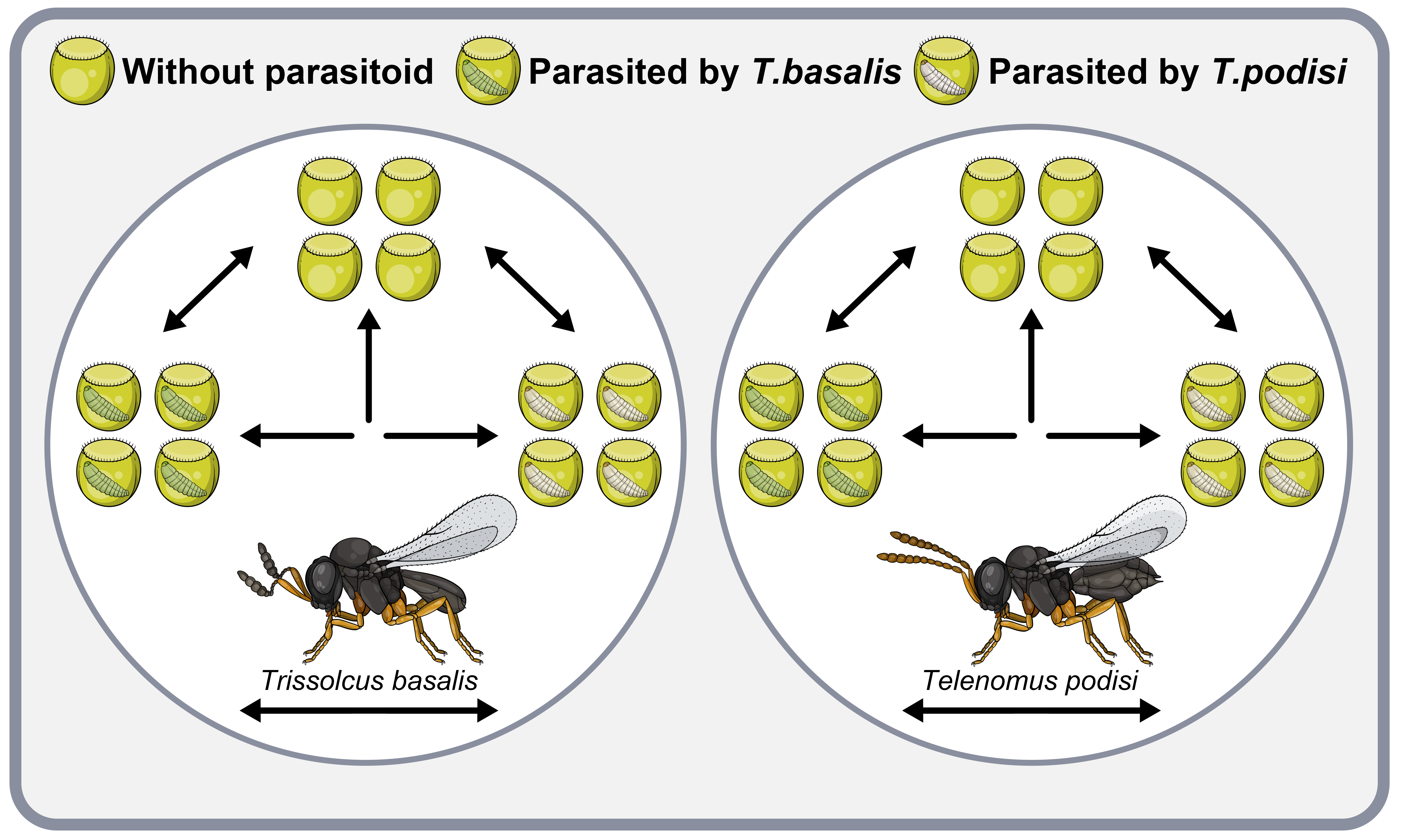}

	\caption{\label{exp_design} Experimental scenarios for quantifying the success of parasitism in the presence of eggs previously parasitised by {\it Trissolcus basalis}, by {\it Telenomus podisi} and not parasitised, in the absence of competition (Adapted from \cite{Bin1993})}
\end{figure}

\begin{figure}[!ht]\centering
	\includegraphics[width=.8\textwidth]{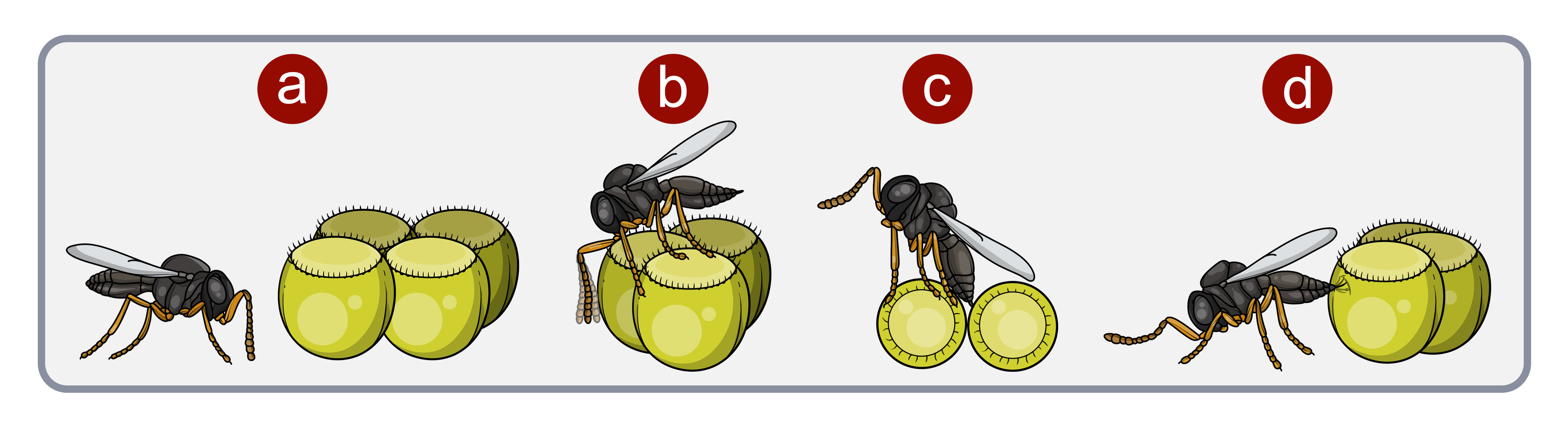}

	\caption{\label{behaviour_Bin} Behaviours exhibited by parasitoids during parasitism  and photos of {\it Telenomus podisi:} a) walking; b) drumming/putting the ovipositor out; c) ovipositing; and d) marking the egg (chemical signalling) (Adapted from \cite{Bin1993})}
\end{figure}

The rearing of the stinkbug and parasitoid species in the laboratory started with insects provided by the Insect Biology Laboratory, Department of Entomology and Acarology, at the University of São Paulo -- USP/ESALQ, in 2021. The rearing of {\it E. heros} was carried out according to a methodology adapted from \cite{mendoza2016}. 
Interactions between parasitoid females of the species {\it T. posidi} or {\it T. basalis} with eggs of the stinkbug \textit{E. heros} took place in experimental arenas, represented by Petri dishes ($15 \times 2$ cm). A total of 12 eggs were made available to a female parasitoid, divided into 3 groups: 4 eggs previously parasitised by females of \textit{T. podisi}, 4 eggs parasitised by \textit{T. basalis}, and 4 unparasitised, as illustrated in Figure \ref{exp_design}. For the observations, the following behaviours were defined and quantified: a) walking; b) drumming; c) ovipositing and d) marking~(Figure \ref{behaviour_Bin}). Each female was observed for 35 minutes, and ten replicates were performed for each parasitoid species.

\newpage

\section{Methods}\label{methodos}

\subsection{Brief review: Stochastic processes and State-space models}\label{review}

The methodological procedures are centred on stochastic processes (Markov chains) and maximum likelihood estimation. As a basis for the central ideas of this work, we present below a review of
concepts related to continuous stochastic processes. For more details, see \cite{karlin2014}.

\definition{A \textit{stochastic process} is a random phenomenon grounded by probabilistic laws that can occur in time or space. It can be denoted by $\{Y_{t}, t \in \tau\}$, where $Y_{t}$ is the random variable associated to the phenomenon indexed by $t$, which takes values in $\tau$, the time (or space) when (where) the process was observed. Depending on the nature of the set $\tau$, the processes can be discrete or continuous. Here, we consider that $Y_{t} \in S$, $S=\{1,2,\ldots, k\}$ is the state space, a set that represents nominal categories~(discrete response), and $\tau=[0,t)$ is a time interval, therefore it is continuous. \label{def1}

According to Definition \ref{def1},  if $y_{0}$ is the initial state of an individual, when it is observed in time $t=0$, it can move to any state in $S$, i.e.:
\[ Y(t): \left\{\begin{array}{lll}
y_{0}, & 0 \leq t < t_{1} \\
y_{1}, & t_{1} \leq t < t_{2} \\
y_{2}, & t_{2} \leq t < t_{3} \\
\vdots & \vdots \\
y_{s}, & t_{s} \leq t < t_{s+1} \\
\end{array} \right. \]
where $y_{0}, y_{1}, \ldots, y_{s} \in S$, but it is not necessary that $y_{t} \neq y_{t+1}$.
Therefore, we assume that in a finite interval, the process has a finite number of jumps for each state $a \in S$. Eventually, it can happen that an individual enters a state and does not leave it, and if this is valid for all individuals, we call it an absorbing state.

\definition{Consider a stochastic process as defined in \ref{def1}, that is, a continuous-time process with discrete $S$, typically also called a ``jump process''. A stochastic process governed by the following law of conditional probability:
\begin{eqnarray}\label{markovprop}
\pi_{a,b}(s,t)&=&P(Y_{(t)}=b \mid Y_{(s)}=y_{a})\\\nonumber
&=& P(Y_{(t)}=b \mid Y_{(t_{0})}=y_{0}, \ldots,
Y_{(t_{n})}=y_{n}, Y_{(s)}=a),
\end{eqnarray}
$\forall s< t \in  \tau$ and $a,b \in S$, is defined as a Markovian process.  The \textit{Markov property} given by Equation (\ref{markovprop}) defines that the probability of a future event, given all history, depends only on the last state. Moreover, \cite{Lara2017b} clarified that these probabilities can be homogeneous over time, and for the continuous case we have
	\begin{eqnarray}\nonumber
			\pi_{ab}(t)=P(Y_{(t+s)}=b \mid Y_{(s)}=a), \hspace{0.2cm} \forall \hspace{0.2cm}  s< t \in  \tau \hspace{0.2cm}  \mbox{and} \hspace{0.2cm}  a,b \in S, 
				\end{eqnarray}
which can be represented, in matricial notation, by $\bm{P}(s,t)=\bm{P}(t)$.

It is also assumed that at each time interval of the process, for every $a \in S$ non-absorbing state, there is a distribution function $F_{a}(t)$ for positive values that characterises the time until the event.
Then, assuming stationarity, i.e, $\pi_{ab}(t)=P(Y_{t+s}=b \mid Y_{s}=a)$, we can show that:
\begin{eqnarray}\label{relation1}
\frac{1-F_{a}(t+s)}{1-F_{a}(s)}=1-F_{a}(t), \hspace{0.3cm} s,t \geq 0,    
\end{eqnarray}
and a distribution that satisfies this condition (\ref{relation1}) is the exponential. Therefore, we may write $F_{a}(t)=1-\exp(-\theta_{a}t)$ if $t \geq 0$, and consequently we have:
\begin{eqnarray}\nonumber
\frac{\partial \pi_{ab}(t)}{\partial t}= -\theta_{a}\pi_{ab}(t)+\theta_{a}\sum_{c \neq a}\pi_{ac}\pi_{cb}(t),
\end{eqnarray}
where $\displaystyle{\theta_{ab}= \left. \frac{\partial \pi_{ab}(t)}{\partial t}\right\rvert_{t=0}}$ for all $a,b \in S$. This defines the transition intensities
\[ \theta_{ab}: \left\{\begin{array}{lll}
-\theta_{a}, & \mbox{if } a=b \\
\theta_{a} \pi_{ab}(t), &  \mbox{if } a \neq b \\
\end{array} \right. . \]
If $a$ is an absorbing state, then $\theta_{a}=0$, but the reciprocal is not true. Therefore, in the continuous case, the matrices
\begin{eqnarray}\nonumber
\bm{P}(t)=\left(\begin{array}{cccc}
\pi_{11}(t) & \pi_{12}(t) &\ldots& \pi_{1k}(t) \\
\pi_{21}(t) & \pi_{22}(t) &\ldots& \pi_{2k}(t) \\
\vdots & \vdots &\ldots& \vdots \\
\pi_{k1}(t) & \pi_{k2}(t) &\ldots& \pi_{kk}(t) \\
\end{array}\right) \mbox{and }
\bm{Q}=\left(\begin{array}{rrrr}
-\theta_{11} & \theta_{12} &\ldots& \theta_{1k} \\
\theta_{21} & -\theta_{22} &\ldots& \theta_{2k} \\
\vdots & \vdots &\ldots& \vdots \\
\theta_{k1} & \theta_{k2} &\ldots& -\theta_{kk} \\
\end{array}\right)
\end{eqnarray}
are jointly important in interpreting response category changes and movement time.

\definition{Let $\bm{x}_{it}=(x_{it1},\ldots,x_{itp})'$ be a vector of covariates associated to a random sample of individuals~($i = 1, 2, \ldots, N)$ with response variable inherent to the stochastic process $Y_{t} \in S = \{1, 2, \ldots, k \}, t \in  \tau$. Assuming that the intensities are the same for all $i$, the \textit{multi-state regression model} is an extension of the generalised linear model~\cite{meira2009}:
\begin{eqnarray}\nonumber
	\theta_{ab}(\cdot) = f\big[\theta_{ab}^{(0)}(\cdot);  \bm{\beta}_{a}^{\top}\bm{x}(t)\big],
\end{eqnarray} 
where $\theta_{ab}^{(0)}(\cdot)$ is the baseline intensity, and $\bm{\beta}_{a}$ is the vector of parameters for each transition $a$. However, the multi-state Cox model~(proportional hazards), that assumes proportionality of the rates of the different transitions, is the most used regression model~\cite{Wreede10}, and is given by
\begin{eqnarray}\label{cox_modelo}
	\theta_{ab}(\bm{x}) = \theta_{ab}^{(0)}\exp[\bm{\beta}_{a}^{\top}\bm{x}(t)],
\end{eqnarray}
and estimated by maximum likelihood.

\subsection{Specific Methods}

For the analysis of the double transitions over time, we consider two non-independent stages. Also, in both stages, we assume the first-order Markov property~(as defined in Section \ref{review}) with a finite number of jumps in the studied time interval. 

The first stage refers to the egg type choice, for which we have the stochastic process indexed by the random variable $\{Y_{1}(t) \in S_{1}, t \in \tau\}$, $\tau=[0,35)$, $S_{1}=\{1,2,3,4\}$,  where  1:~unparasitised eggs; 2:~eggs parasitised by {\it T. podisi}; 3:~eggs parasitised by {\it T. basalis }  and 4:~no choice.  For the second stage of the process, i.e. behaviour given choice, we include an additional category to consider the different insect behaviours after choice made in the first stage. We have therefore the conditional stochastic process $\{[Y_{2} \mid Y_{1} ](t) \in S_{2},  t \in \tau\}, S_{2}=\{1,2,3,4\}$, in which they represent  1:~marking; 2:~ovipositing 3:~drumming  and 4:~ returning to set $S_{1}$, hereafter named as ``other''. The transition scheme is represented in Figure \ref{esquema_double}, in which we can see that there are no absorbing states.

\begin{figure}[!ht]\centering
	\includegraphics[width=1\textwidth]{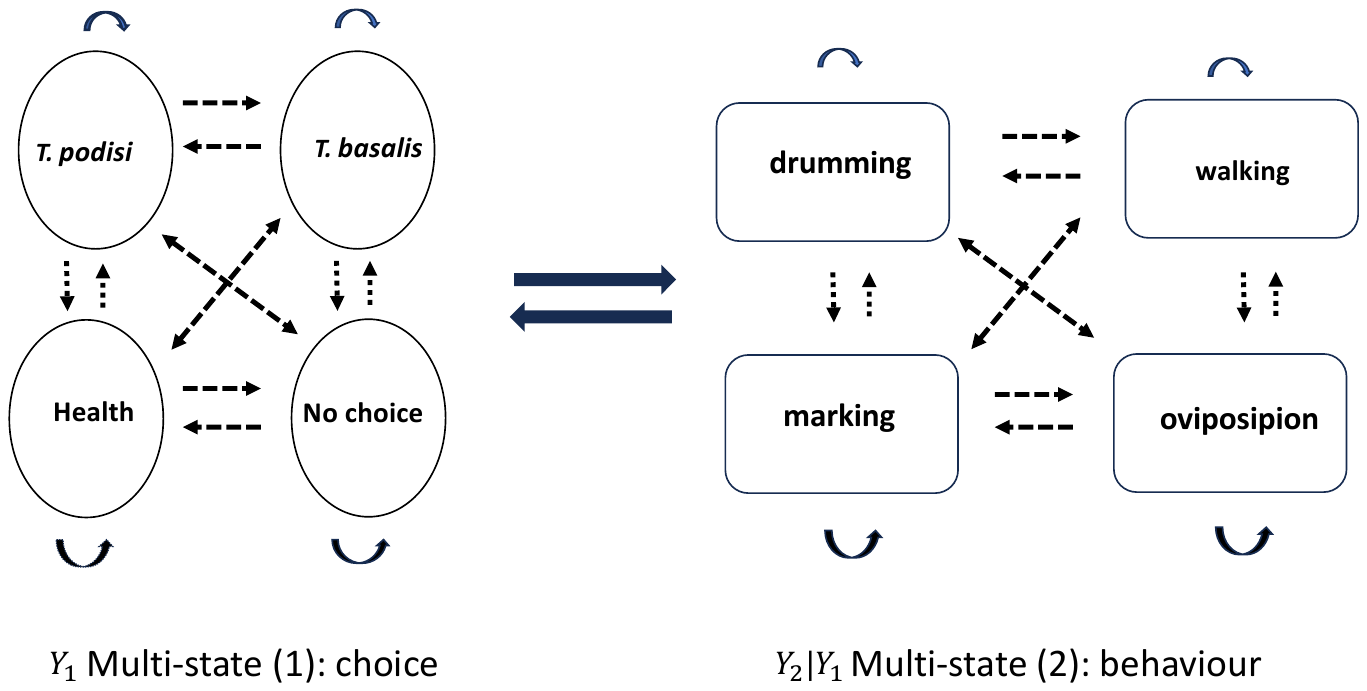}
 \vspace{-2.5cm}
	\caption{\label{esquema_double} Double transitions scheme: choice in the first stage and behaviour given choice.}
\end{figure}

Assuming the processes are stationary, we may consider the transition intensities matrix 
\begin{eqnarray} \nonumber
\scriptsize{\bm{Q}=\left(\begin{array}{cccc}
		-(\theta_{1}+\theta_{2}+\theta_{3}) & \theta_{1}&\theta_{2}& \theta_{3} \\
		\theta_{4} & 	-(\theta_{4}+\theta_{5}+\theta_{6})  &\theta_{5}& \theta_{6} \\
		\theta_{7}& \theta_{8} &	-(\theta_{7}+\theta_{8}+\theta_{9}) & \theta_{9} \\
		\theta_{10} & \theta_{11}&\theta_{12}& 	-(\theta_{10}+\theta_{11}+\theta_{12})  \\
	\end{array}\right)	}	
\end{eqnarray}		
for both stages, assuming a stochastic double random walk for the parasitoids. Here, the parameters are functionally related to the effects of time and species~(treatment factor) through the Markovian Cox-model (equation \ref{cox_modelo}):
\begin{eqnarray}\label{cox_model}
	\theta_{ab} = \theta_{ab}^{(0)}\exp\{\bm{\beta}_{a}^{\top}\mbox{species}(t)\}, \hspace{0.5cm} \forall \hspace{0.2cm} a, b  \in S.
	\label{model1}
\end{eqnarray}
The null model for both processes considers only the time effect, while the full model incorporates the treatment effect as described by equation \ref{cox_model} above. Furthermore, the parameters are estimated by maximum likelihood via an iterative algorithm, whose initial values $\theta_{ab}^{(0)}$ are calculated based on the observed transition frequencies and the time spent in each state:
\begin{eqnarray}\nonumber
\theta_{ab}^{(0)}=   \left\{ \begin{array}{rr}
			\displaystyle{-\frac{n_{a.}}{T_{a.}}},  & \mbox{if } k = k'    \\\nonumber
   \\
			\displaystyle{\frac{n_{ab}}{T_{a.}}},  & \mbox{if }  k \neq k'  \end{array} \right. \\ \nonumber   
\end{eqnarray}
where $n_{a.}$ denotes the number of insects that were in state $a$ at the previous time, $n_{ab}$ are the transition frequencies from $a$, and $T_{a.}$ is the total time spent at the previous state $a$.

To differentiate the two stages, we denote the intensities matrices by $\bf{Q}_{Y_{1}}$ and $\bf{Q}_{Y_{2} \mid Y_{1}}$, and the transition probability matrices by $\bf{P}_{Y_{1}}(t)$ and  $\bf{P}_{Y_{2} \mid Y_{1}}(t)$. The transition probabilities, and the time spent in each state, are estimated by the invariance principle of the maximum likelihood estimators. Finally, we employ likelihood-ratio ($\Lambda$) tests to assess the significance of the treatment effect. All analyses were carried out in R \cite{R22} using packages \texttt{survival} \cite{Therneau} and \texttt{msm} \cite{Jackson2009}.

\section{Results }\label{results}

We begin showing the contingency tables:  treatment versus egg choice (Table\ref{tab1}) and  treatment versus behaviour unconditional to egg choice (Table \ref{tab2}).  Higher frequencies were observed for choosing unparasitesed eggs and, in relation to behaviour, marking had higher frequencies for both species.  
According to the $\chi^2$ test, we verify an association between treatment (parasitoid species) and egg choice ($p <0.001$), and a non-association with respect to the treatments and  behaviours ($p=0.9596$). 

\begin{table}[!ht]\centering
	\caption{Number of choices observed according to species, obtained in the parasitism data, according with the experiment developed by \cite{Bon2021}. \label{tab1}}
		\begin{tabular}{lrr}\hline
\multirow{2}{*}{Choices}& \multicolumn{2}{c}{Treatments} \\
& {\it Trissolcus basalis} & {\it Telenomus posidi} \\ \hline
unparasitised eggs                             & 150  &  125   \\
eggs parasitised by {\it T. podisi}            & 97   &    9\\
eggs parasitised by {\it T. basalis }          & 35   &   124 \\
no choice                                      & 65   &  60   \\
			\hline
	\end{tabular}
\end{table}

\begin{table}[!ht]\centering
	\caption{Observed Number behaviour unconditional to egg choice  according to species, obtained in the parasitism data, according with the experiment developed by \cite{Bon2021}. \label{tab2}}
		\begin{tabular}{lrr}\hline
\multirow{2}{*}{Behaviours}& \multicolumn{2}{c}{Treatments} \\
& {\it Trissolcus basalis} & {\it Telenomus posidi} \\ \hline
marking                                        & 126  &  121   \\
ovipositing                                    & 64  &    56\\
drumming                                       & 92  &   81 \\
returning to set $S_{1}$                      & 65   &  60   \\
			\hline
	\end{tabular}
\end{table}

Specifically regarding  transitions over time, a  total of 641 were observed, with a  minimum of 12 and a maximum of 40 transitions (27.7 on average) both for choosing eggs and for behaviours, without taking into account the treatment structure. Evidently, the frequencies observed per treatment and transitions are merely exploratory techniques, the significance or otherwise of the effects depends on the time and intensities of transitions.

\subsection{First transition -- choice}

Next, we model egg choice (i.e., the $Y_{1}$ process) using the Cox model. The treatment effect was significant ($\Lambda=68.62$; d.f. = 7; $p< 0.001$). For this first stage, the estimated transition intensities matrices were:
\begin{eqnarray} \nonumber
\bm{\hat{Q}}_{Y_{1}}(T.~basalis)=\left(\begin{array}{rrrr}
			-0.339 & 0.000 & 0.000  & 0.339 \\
			0.000 & -0.202 & 0.001  & 0.200 \\
			0.000 & 0.000 &-0.690  &0.690 \\
			1.607  & 0.616  & 0.650  &-2.874 \\
		\end{array}\right)	
\end{eqnarray}		
and
\begin{eqnarray} \nonumber
	\bm{\hat{Q}}_{Y_{1}}(T.~podisi)=\left(\begin{array}{rrrr}
		-0.182 & 0.000 & 0.000 & 0.182 \\
		0.000 &-8.403 & 0.003 & 8.401 \\
		0.000 & 0.000 &-0.308 & 0.308 \\
		1.130 & 1.805  &1.720 & -4.655 \\
	\end{array}\right)	,
\end{eqnarray}		
showing differences between transition rates, especially in the diagonals, for which higher values of transition intensity imply a longer time to exit the state. Thus, considering a recognition of previously parasitised eggs, the exit rate from the ``eggs previously parasitised by {\it T. podisi}'' state is lower for {\it T. podisi} ($-8.403$) when compared to {\it T. basalis} ($-0.202$).

From the intensities matrices, we may obtain the mean times and respective confidence intervals for the choices per each treatment~(Figure \ref{IC_escolha}). We observe that {\it T. podisi} does not spend time choosing the eggs already parasitised by conspecifics, but this is not true when eggs had been parasitised by its competitor \textit{T. basalis}. Moreover, we emphasise that null transition intensities do not imply null transition probabilities, but rather that the exit from the state does not occur instantaneously. Using the estimated intensities it is possible to obtain the transition probabilities matrices~(see Figure \ref{probs_escolha}). We see that \textit{T. basalis} is less selective when ovipositing, with higher transition probabilities to superparasitism and/or multiparasitism behaviours.

\begin{figure}[!ht]\centering
	\includegraphics[width=\textwidth]{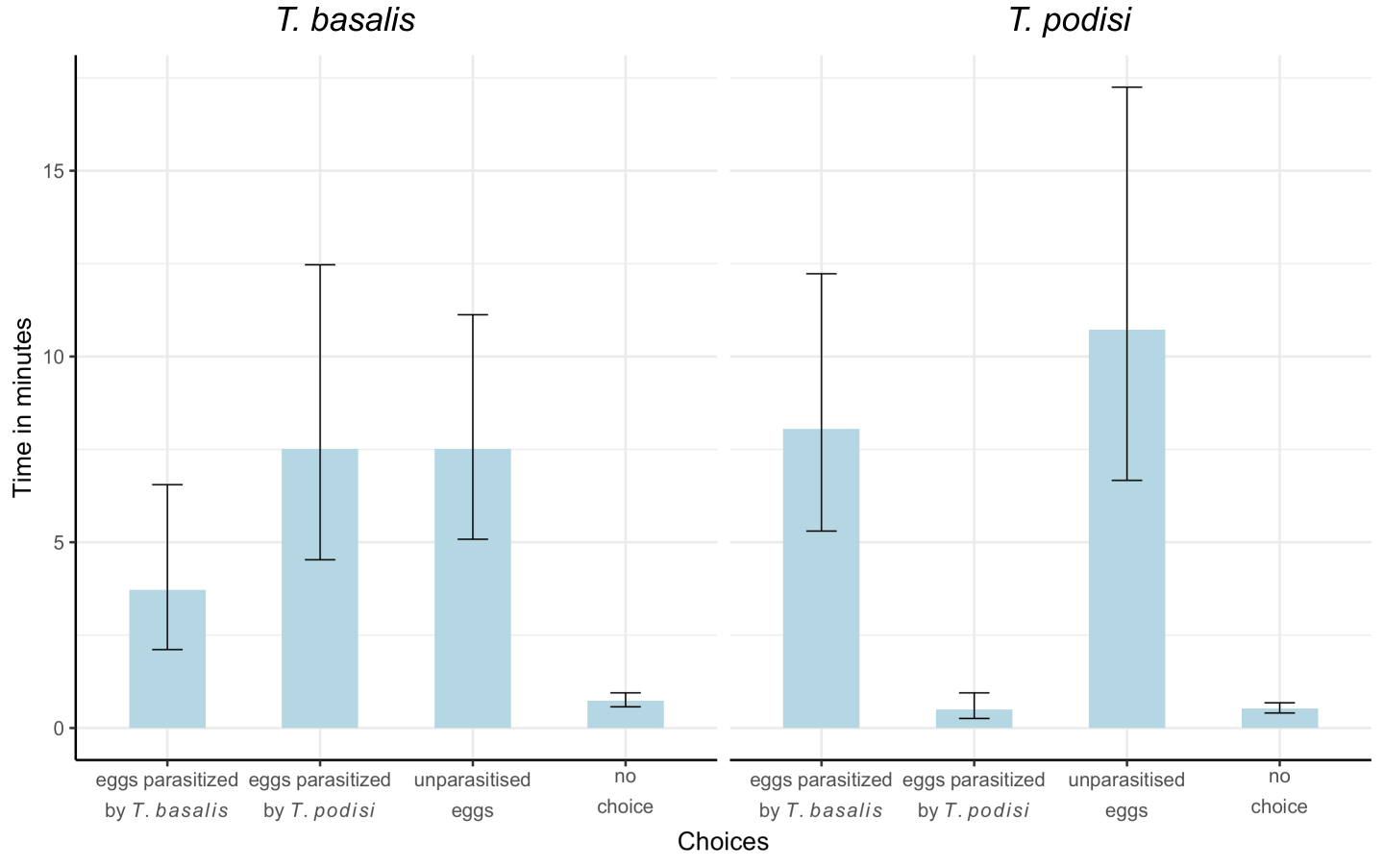}
	\caption{\label{IC_escolha} Means and confidence intervals of time spent in each choice, for both treatments (parasitoid species).}
\end{figure}

\begin{figure}[!ht]\centering
	\includegraphics[width=1.05\textwidth]{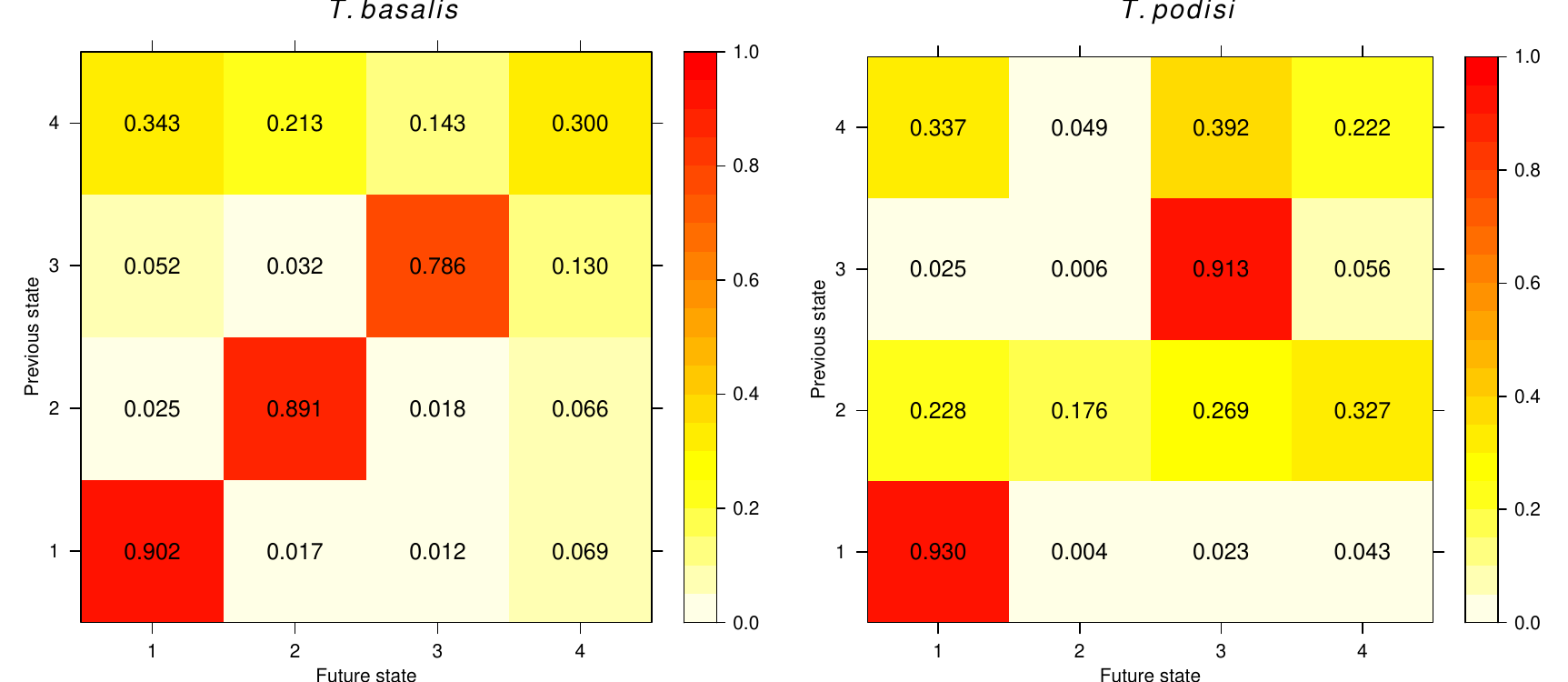}
	\caption{\label{probs_escolha} Estimated probabilities for parasitoid choices (first step), considering 1: unparasitised eggs; 2: eggs parasitised by {\it T. podisi}; 3: eggs parasitised by {\it T. basalis }  and 4: no choice.}
\end{figure}

\subsection{Second transition - Behaviour given choice}
Now considering the behaviours given a choice, the treatment effect was also significant for $Y_{2} \mid Y_{1}:$ unparasitised eggs ($\Lambda=33.20$; d.f. = 9; $p<0.001$),  for $Y_{2} \mid Y_{1}:$  eggs previously parasitised by {\it T. basalis} ($\Lambda=19.33$; d.f. = 9; $p=0.022$), and for $Y_{2} \mid Y_{1}:$ eggs previously parasitised by {\it T. podisi} ($\Lambda=55.53$; d.f. = 8; $p< 0.001$).

Regarding the estimated values, first, conditionally on the choice of unparasitised eggs, a greater difference was observed in the transition intensities associated with ovipositing (second line of transition intensities matrices), which indicates that the transition intensity for {\it T. podisi} is higher than for {\it T. basalis}, except when going back to stage 1 (unparasitised eggs):
	\begin{eqnarray} \nonumber
		\bm{\hat{Q}}_{Y_{2} \mid Y_{1}=1}(T.~basalis)=\left(\begin{array}{rrrr}
			-0.961 & 0.107 & 0.748 & 0.107 \\
			3.320 & -3.566 & 0.123  & 0.1229 \\
			0.000 & 0.197 &-0.299 &0.102\\
			0.000  & 0.000 & 1.425  &-1.425\\
		\end{array}\right)	
	\end{eqnarray}
and
	\begin{eqnarray} \nonumber
		\bm{\hat{Q}}_{Y_{2} \mid Y_{1}=1}(T.~podisi)=\left(\begin{array}{rrrr}
			-0.916 & 0.095 & 0.790 & 0.032 \\
			0.755 &-1.006 & 0.226 & 0.0252 \\
			0.000 & 0.195 &-0.273 & 0.077 \\
			0.000 & 0.000 &2.119& -2.119 \\
		\end{array}\right) .	
	\end{eqnarray}

The estimated transition probabilities matrices are presented in Figure \ref{probs_sadios}. Despite the apparent homogeneity, in practice, it is noted that, in general, the probabilities of transitions to ovipositing~(state 2) and drumming~(state 3) are higher for {\it T. podisi}. In summary, when choosing unparasitised eggs, the {\it T. podisi} is faster than \textit{T. basalis}, and presents a higher ovipositing rate.

\begin{figure}[!ht]\centering
	\includegraphics[width=1.05\textwidth]{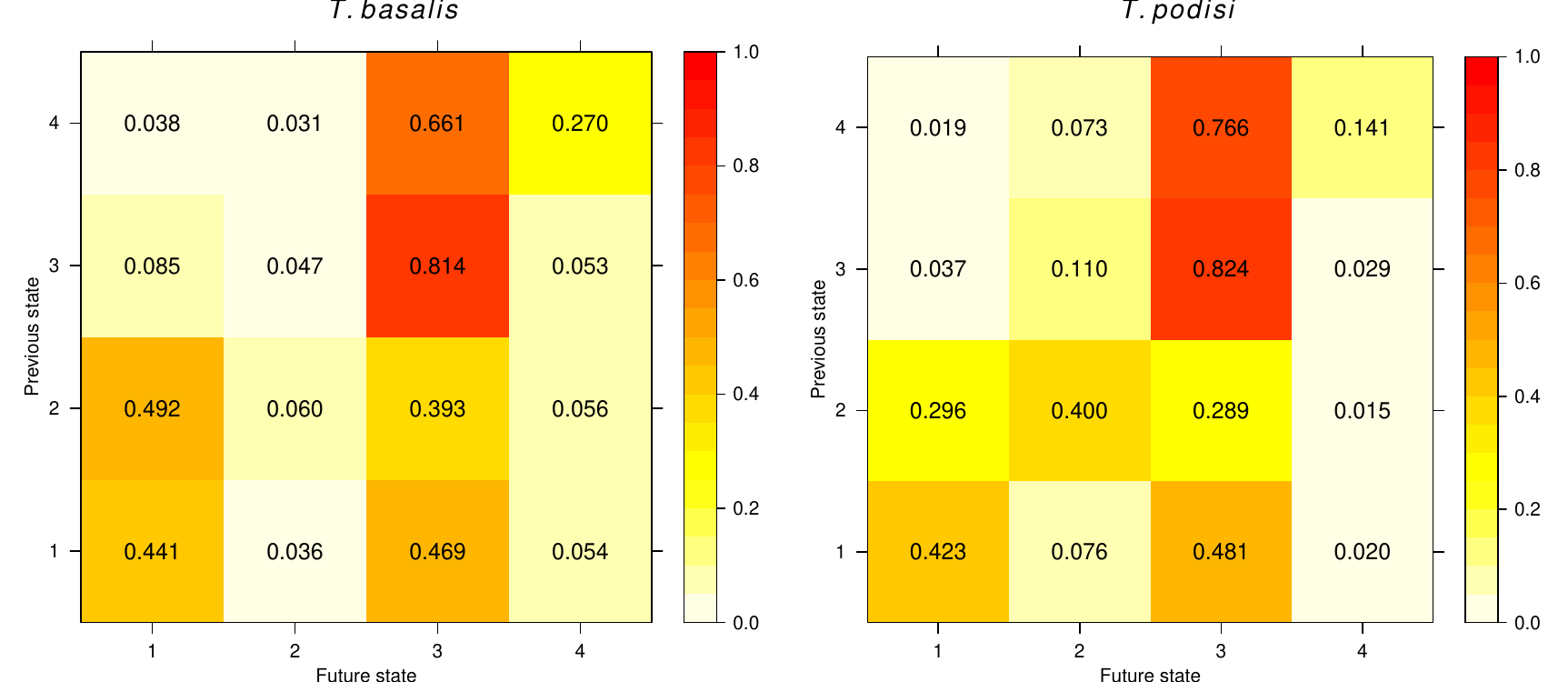}
	\caption{\label{probs_sadios} Estimated transition probabilities conditional on choosing unparasitised eggs for parasitoid behaviours 1: marking, 2: ovipositing, 3: drumming, and 4: others.} 
\end{figure}

Now, considering the behaviour conditional on the choice of eggs previously parasitised by {\it T. podisi} and, firstly, analyzing the estimates of transition intensities, there is a considerable difference in point values when comparing the matrices of the two treatments~(species):
	\begin{eqnarray} \nonumber
		\bm{\hat{Q}}_{Y_{2} \mid Y_{1}=2}(T.~basalis)=\left(\begin{array}{rrrr}
			-1.405 & 0.000 & 1.003 & 0.401 \\
			2.745 & -3.137 & 0.392  & 0.000 \\
			0.000 & 0.076 &-0.170 &0.095\\
			0.000  & 0.105 & 1.368  &-1.474\\
		\end{array}\right)	
	\end{eqnarray}		
and	
	\begin{eqnarray} \nonumber
		\bm{\hat{Q}}_{Y_{2} \mid Y_{1}=2}(T.~podisi)=\left(\begin{array}{rrrr}
			-0.695 & 0.000 & 0.618 & 0.0772 \\
			0.990&-1.247 & 0.257 & 0.000 \\
			0.006& 0.200 &-0.316 & 0.111 \\
			0.000 & 0.000 &1.733& -1.733 \\
		\end{array}\right)	.
	\end{eqnarray}
For example, regarding the behaviour of marking eggs (first row of the matrices), {\it T. podisi} presented a lower intensity, signalling that the parasitoids make the transition through this state quickly, which indicates that they accepted the eggs for successful oviposition. The estimated transition probabilities matrices are presented in Figure \ref{probs_podisi}.

\begin{figure}[!ht]\centering
	\includegraphics[width=1.05\textwidth]{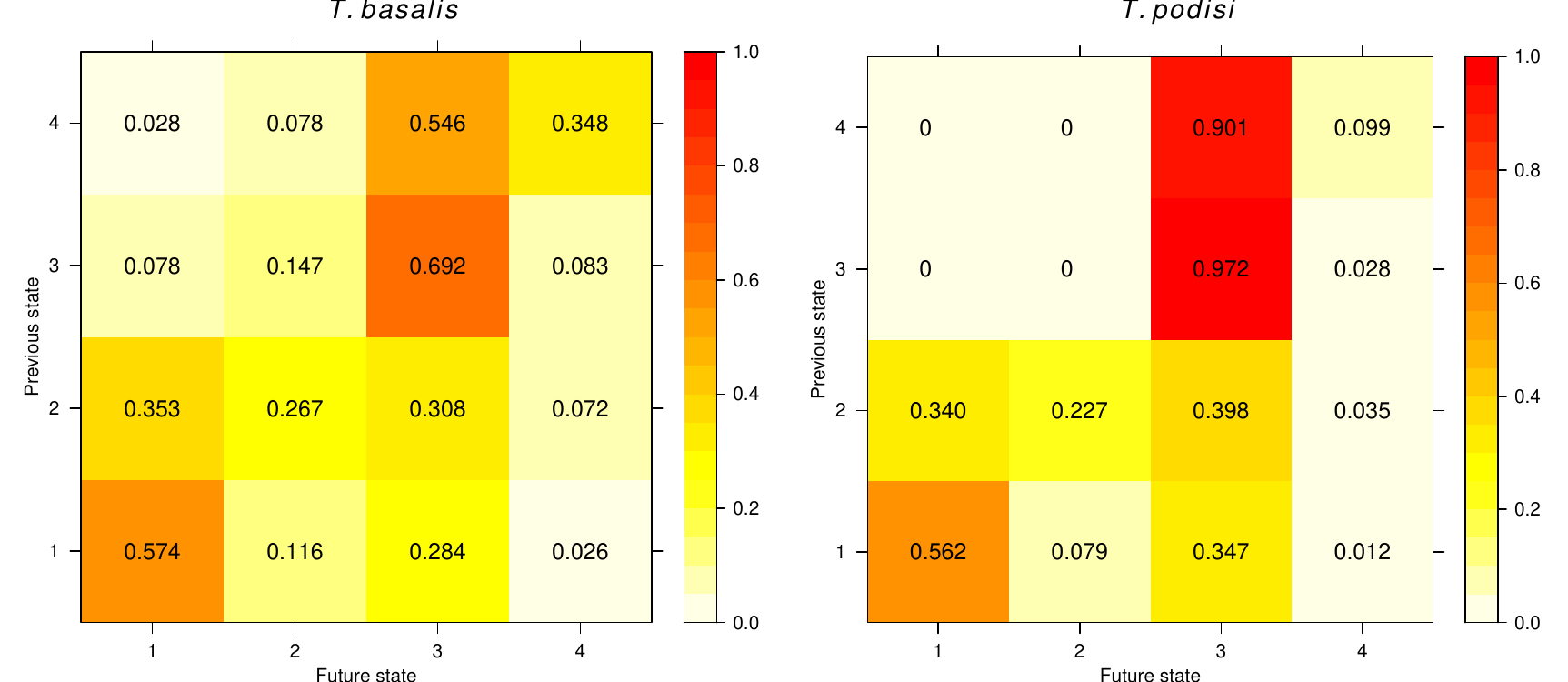}
	\caption{\label{probs_podisi} Estimated transition probabilities conditional on choosing eggs previously parasitised by \textit{T. podisi} for parasitoid behaviours 1: marking, 2: ovipositing, 3: drumming, and 4: others.}
\end{figure}

Additionally, evaluating the estimates of the transition probabilities~(Figure \ref{probs_podisi}), there is a lower probability of oviposition of this species compared to {\it T. basalis}~(note that there are zero probabilities of transition from other states and drumming to ovipositing), possibly due to the previous recognition of previous conspecific parasitism, and on the other hand, higher probabilities of transition to drumming, which is an intermediate behaviour for state change.

Considering the behaviours given a choice of eggs previously parasitised by {\it T. basalis}, the transition intensities are similar for marking and ovipositing, but not for drumming and returning to stage 1 (others):
\begin{eqnarray} \nonumber
		\bm{\hat{Q}}_{Y_{2} \mid Y_{1}=3}(T.~basalis)=\left(\begin{array}{rrrr}
			-0.659 & 0.231& 0.428 & 0.000 \\
			0.991 & -1.6351 & 0.446  & 0.198 \\
			0.000 & 0.370 &-0.521&0.151\\
			0.000  & 0.000 & 1.180  &-1.1803\\
		\end{array}\right)	
	\end{eqnarray}
and	
	\begin{eqnarray} \nonumber
		\bm{\hat{Q}}_{Y_{2} \mid Y_{1}=3}(T.~podisi)=\left(\begin{array}{rrrr}
	-0.659 & 0.231& 0.428 & 0.000 \\
    0.991 & -1.635 & 0.446  & 0.198 \\
    0.000 & 0.000&-0.0803&0.0803\\
    0.000  & 0.000 & 2.571  &-2.571\\
		\end{array}\right)	.
	\end{eqnarray}
The drumming rates are higher for {\it T. basalis}, while {\it T.podisi}  presented lower intensity rates associated with `other' behaviours. Stochastically, the transition probabilities for these states also showed significant differences, as illustrated in Figure \ref{probs_basalis}.

\begin{figure}[!ht]\centering
	\includegraphics[width=1.05\textwidth]{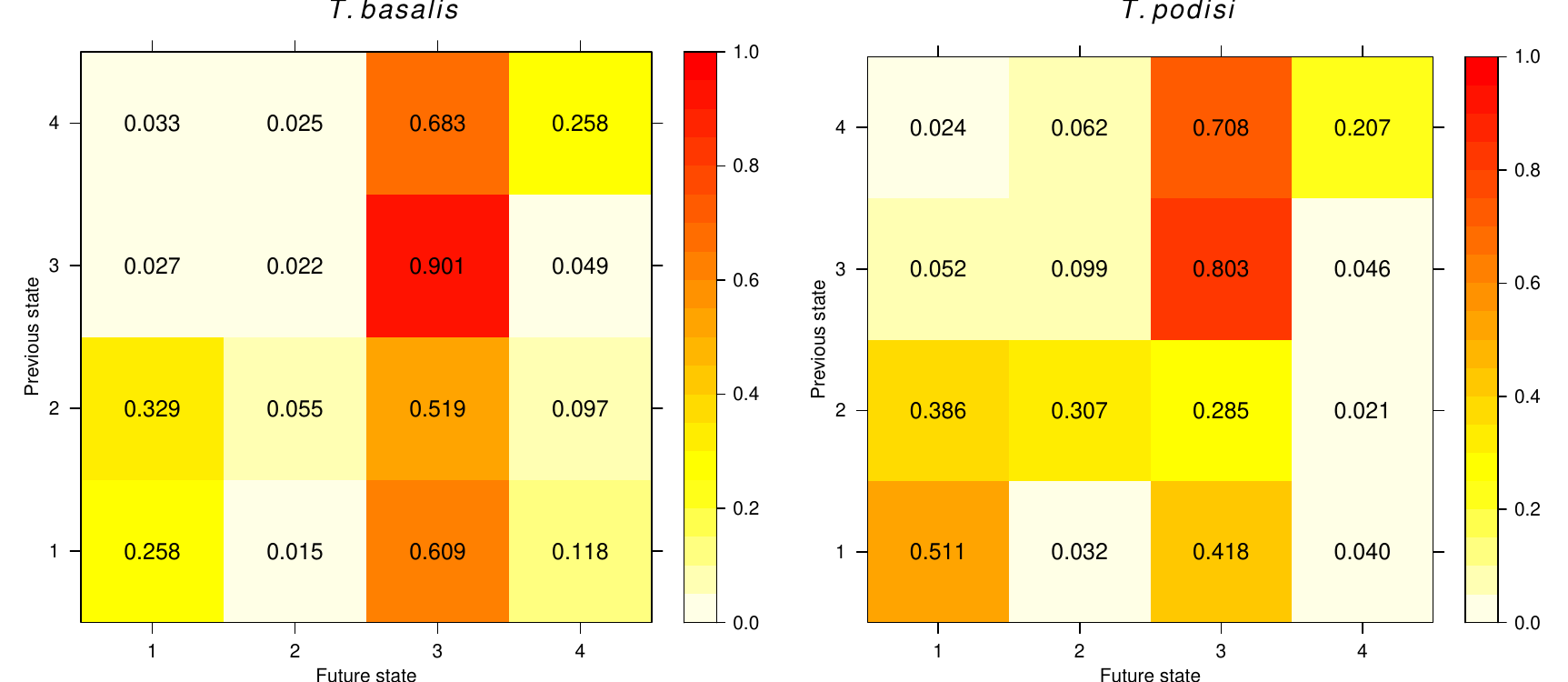}
	\caption{\label{probs_basalis} Estimated transition probabilities conditional on choosing eggs previously parasitised by \textit{T. basalis} for parasitoid behaviours 1: marking, 2: ovipositing, 3: drumming, and 4: others.}
\end{figure}

Finally, the times and respective 95\% confidence intervals of the transition intensity times associated with each behaviour, given the a choice of type of eggs, for each species are presented in the Figure \ref{IC_T1}. These figures reiterate the rapid action associated with the marking and ovipositing behaviours of {\it T. podisi} when compared to {\it T. basalis} for eggs previously parasitised by conspecifics.

\begin{figure}[!ht]\centering
	\includegraphics[width=\textwidth]{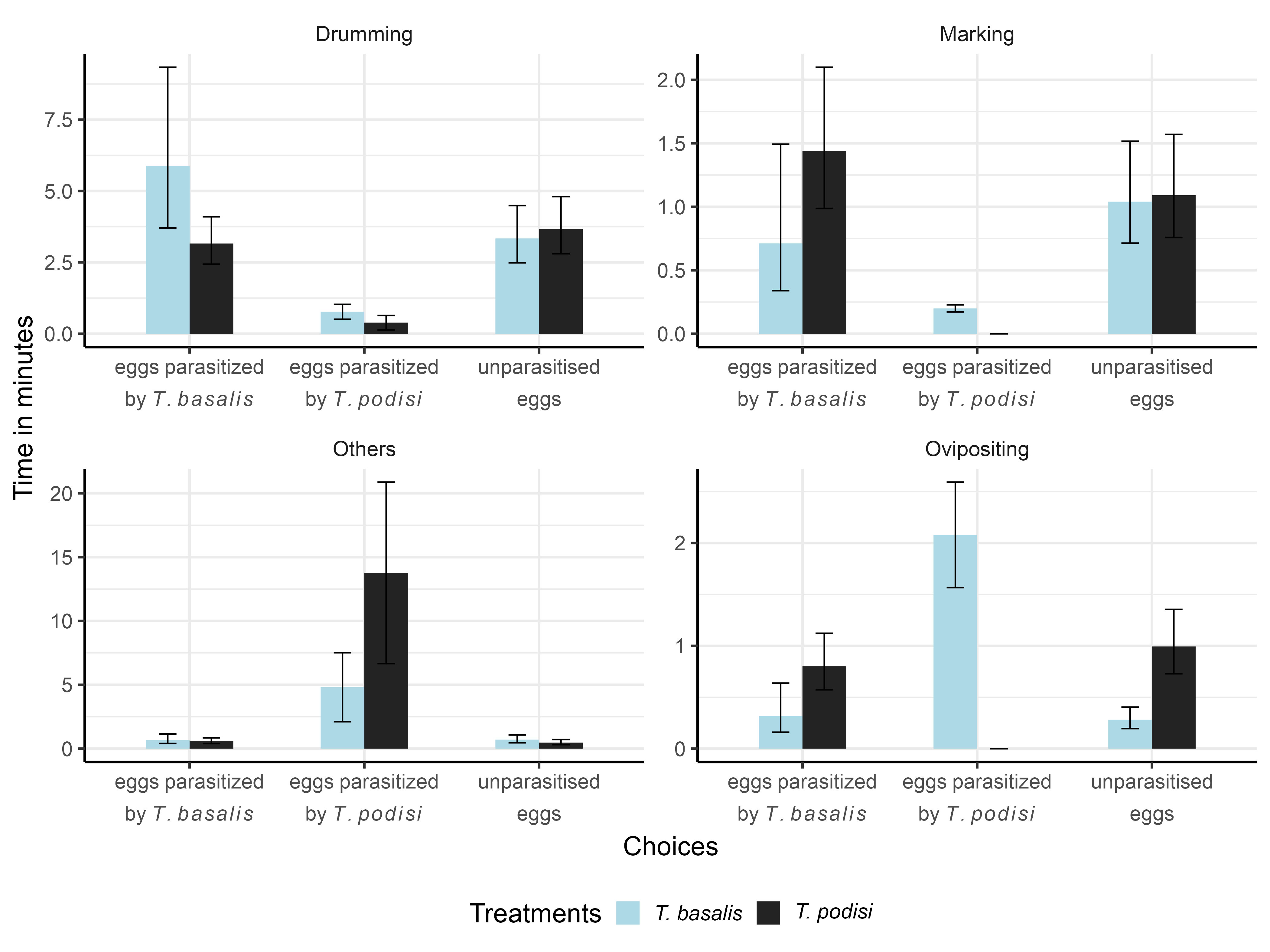}
	\caption{\label{IC_T1} Means and 95\% confidence interval for time spent in each behaviour given choice, related to transition intensities considering the treatments {\it T. basalis} {\it T. podisi} and species.}
\end{figure}

\newpage

\section{Discussion }\label{discussion}

The ability to recognize physical or chemical marks left on the host after oviposition is considered a natural tendency of parasitoid species to avoid superparasitism/multiparasitism, which can lead the parasitoid offspring to be forced into a lethal competition \cite{godfray1994}. In this study, both, {\it T. basalis} and {\it T. podisi} females displayed a comparable sequence of host handling behaviours: drumming, oviposition, and host marking. A major difference, however, was observed in how long was displayed the oviposition behaviour on eggs previously parasitised. Females of {\it T. basalis} expend more time ovipositing in previously parasitised eggs by {\it T. podisi}, exhibiting a higher tendency to multiparasitism.  {\it Trissolcus basalis} females also oviposit in host eggs previously parasitised by conspecifics, leading to superparasitism, and increasing the chances of competition between siblings, since only one egg typically can successfully develop into adulthood. 

On the other hand, {\it T. podisi} females could avoid the eggs previously parasitised by conspecifics and, additionally, could find unparasitised eggs and oviposit on them faster than {\it T. basalis}. Because both {\it T. podisi} and {\it T. basalis} were exposed to the same parasitoid and host densities in our experiment, we can assume that {\it T. basalis} has a stronger natural tendency to self-superparasitise and multiparasitise than {\it T. podisi}. In terms of host foraging, {\it T. podisi} exhibited higher search efficiency, showing a high capability to find healthy hosts and avoid super and multiparasitism.  

The occurrence of host marking is a reliable indicator of successful oviposition in scelionids \cite{wiedemann2003}. Considering the marking behaviour, our results showed that {\it T. basalis} females expend more time oviposit in groups of previously parasitised host eggs available. In agreement with the previous study \cite{bon2022}, our results found that {\it T. basalis} can reduce the host population. However, in a condition of very high intra- and interspecific competition, the reduction of the host population cannot result in an increase of the parasitoid populations for subsequent generations. 

\section{Conclusions} \label{conclusion}

Biological pest control is a sustainable practice that benefits food production and health. Despite all the biological and environmental appeal for the development of these studies, there is also a need for adequate statistical methodologies to confirm the scientific hypotheses. Interactions between species, as well as changes in behaviour over time require specific methods of analysis to estimate the biological control efficiency of a species, and models for categorical longitudinal data are very useful in this context.

In this work, we presented the problem of the soybean pest \textit{Euschistus heros} and two potential agents for natural control in the field. As a statistical contribution, we developed an extension of multi-state models to compare two parasitoid species by evaluating their behaviours over time. These models allow not only to describe behavioural actions but also the intensity with which they occur. In this context, the method  validated the experimental assumption that the species {\it T. podisi} avoids intra-specific competition by being more efficient in recognising and avoiding previous conspecific parasitism. In the applied sense, the results can contribute to improving the parasitoid release strategies in the field and optimise the mass-rearing production. Moreover, the proposed statistical method used can also be a contribution to potential researchers studying insect behaviour.
Although in this work the method has proved to be effective, for future studies there is a need to consider sub-intervals of time, for which we can allow different transition rates, since they may not be homogeneous over time.

\section*{Acknowledgements}

This work had financial support from the Brazilian Foundation, Coordenação de ``Coordena\c{c}\~{a}o de Aperfei\c{c}oamento de Pessoal de N\'{i}vel Superior'' (CAPES)  process number $88887.716582/2022-00$. This publication has emanated from research conducted with the financial support of Science Foundation Ireland under Grant number 18/CRT/6. The authors are grateful to John Hinde for valuable suggestions that helped improve the manuscript.

\bibliographystyle{tfs}
\bibliography{interacttfssample}

\begin{thebibliography}{10}
\providecommand{\MR}{\relax\unskip\space MR }
\providecommand{\url}[1]{\normalfont{#1}}
\providecommand{\urlprefix}{Available at }

\bibitem{Agresti2010}
A. Agresti, \emph{Analysis of ordinal categorical data}, Vol. 656, John Wiley
  \& Sons, 2010.

\bibitem{Bin1993}
F. Bin, S. Vinson, M. Strand, S. Colazza, and W. Jones~Jr, \emph{Source of an
  egg kairomone for trissolcus basalis, a parasitoid of nezara viridula},
  Physiological Entomology 18 (1993), pp. 7--15.

\bibitem{Bon2021}
V.J. Bon, \emph{Efeito da intera{\c{c}}{\~a}o de trissolcus basalis e telenomus
  podisi (hymenoptera: Scelionidae) na efetividade do controle biol{\'o}gico de
  euschistus heros (hemiptera: Pentatomidae)}, Master Thesis  (2021).

\bibitem{bon2022}
V.J. Bon, R. de  Andrade~Moral, and C. Reigada, \emph{Influence of intra-and
  inter-specific competition between egg parasitoids on the effectiveness of
  biological control of euschistus heros (hemiptera: Pentatomidae)}, Biological
  Control 170 (2022), p. 104903.

\bibitem{bs1994}
C.F. BS, \emph{Temperature-effect on the biology and reproductive performance
  of the egg parasitoid trissolcus basalis (woll.)}, Anais da Sociedade
  Entomol{\'o}gica do Brasil 23 (1994), pp. 399--408.

\bibitem{camarozano2021}
C.T. Camarozano, A. Coelho~Jr, R.B.Q.d. Silva, and J.R.P. Parra, \emph{Can
  trichogramma atopovirilia oatman \& platner replaces trichogramma galloi
  zucchi for diatraea saccharalis (fabricius) control?}, Scientia Agricola 79
  (2021).

\bibitem{correa1993}
B.S. CORREA-FERREIRA, \emph{Utiliza{\c{c}}{\~a}o do parasit{\'o}ide de ovos
  trissolcus basalis (wollaston) no controle de percevejos da soja.}, PhD
  Thesis  (1993).

\bibitem{Correia1991}
B. Corrêa-Ferreira, \emph{Parasitóide de ovos: incidência natural, biologia
  e efeito sobre a população de percevejos da soja}, Phd Thesis 3 (1979), pp.
  36--39.

\bibitem{cusumano2016}
A. Cusumano, E. Peri, and S. Colazza, \emph{Interspecific
  competition/facilitation among insect parasitoids}, Current opinion in insect
  science 14 (2016), pp. 12--16.

\bibitem{de2012}
A. de  Freitas~Bueno, D.R. Sosa-G{\'o}mez, B.S. Corr{\^e}a-Ferreira, F.
  Moscardi, and R.C.O. de  Freitas~Bueno, \emph{Inimigos naturais das pragas da
  soja}, Soja: manejo integrado de insetos e outros artr{\'o}podes-praga.
  Bras{\'\i}lia, Brasil, EMBRAPA  (2012), pp. 493--522.

\bibitem{Lara2017}
I. de  Lara, J. Hinde, A. De~Castro, and I. Da~Silva, \emph{A proportional odds
  transition model for ordinal responses with an application to pig behaviour},
  Journal of Applied Statistics 44 (2017), pp. 1031--1046.

\bibitem{Lara2017b}
I.A.R. de  Lara, J. Hinde, and C.A. Taconeli, \emph{An alternative method for
  evaluating stationarity in transition models}, Journal of Statistical
  Computation and Simulation 87 (2017), pp. 2962--2980.

\bibitem{Paula2023}
L.V.T. de  Paula, I.A.R. de  Lara, C. Reigada, and V.J. Bon, \emph{Multistate
  model in the behavioral study of the parasitoid telenomus podisi for
  biological soybean control}, Brazilian Journal of Biometrics 41 (2023), pp.
  70--82.

\bibitem{Wreede10}
L.C. De~Wreede, M. Fiocco, and H. Putter, \emph{The mstate package for
  estimation and prediction in non-and semi-parametric multi-state and
  competing risks models}, Computer methods and programs in biomedicine 99
  (2010), pp. 261--274.

\bibitem{Diggle2002}
P. Diggle, P.J. Diggle, P. Heagerty, K.Y. Liang, P.J. Heagerty, S. Zeger,
  \emph{et~al.}, \emph{Analysis of longitudinal data}, Oxford University Press,
  2002.

\bibitem{follett2000}
P.A. Follett, J. Duan, R.H. Messing, and V.P. Jones, \emph{Parasitoid drift
  after biological control introductions: re-examining pandora's box}, American
  Entomologist 46 (2000), pp. 82--94.

\bibitem{gallo2002}
D. Gallo, O.N. Nakano, S. Silveira~Neto, R.P.L.C. Carvalho, G.C.D.D. Batista,
  E. Berti~Filho, J.R.P.L. Parra, R.A. Zuchi, and S. Bat, \emph{Entomologia
  agr{\'\i}cola}, in \emph{Entomologia agr{\'\i}cola}, FEALQ,  2002, pp.
  920--920.

\bibitem{godfray1994}
H.C.J. Godfray, \emph{Parasitoids: behavioral and evolutionary ecology},
  Vol.~67, Princeton University Press, 1994.

\bibitem{harvey2013}
J.A. Harvey, E.H. Poelman, and T. Tanaka, \emph{Intrinsic inter-and
  intraspecific competition in parasitoid wasps}, Annual review of entomology
  58 (2013), pp. 333--351.

\bibitem{Jackson2009}
 {Jackson, C}, \emph{msm: multi-state Markov and hidden Markov models in
  continuous time (version 0.9.3)}, R Foundation for Statistical Computing,
  Vienna, Austria (2009). \urlprefix\url{https://www.R-project.org/}.

\bibitem{karlin2014}
S. Karlin, \emph{A first course in stochastic processes}, Academic press, 2014.

\bibitem{Lara2020}
I.A. Lara, R.A. Moral, C.A. Taconeli, C. Reigada, and J. Hinde, \emph{A
  generalized transition model for grouped longitudinal categorical data},
  Biometrical Journal 62 (2020), pp. 1837--1858.

\bibitem{meira2009}
L. Meira-Machado, J. de  U{\~n}a-{\'A}lvarez, C. Cadarso-Su{\'a}rez, and P.K.
  Andersen, \emph{Multi-state models for the analysis of time-to-event data},
  Statistical methods in medical research 18 (2009), pp. 195--222.

\bibitem{mendoza2016}
A.C. Mendoza, A.C. da  Rocha, and J.R. Parra, \emph{Lyophilized artificial diet
  for rearing the neotropical euschistus heros (hemiptera: Pentatomidae)},
  Journal of Insect Science 16 (2016), p.~41.

\bibitem{queiroz2018}
A. Queiroz, E. Taguti, A. Bueno, M. Grande, and C. Costa, \emph{Host
  preferences of telenomus podisi (hymenoptera: Scelionidae): parasitism on
  eggs of dichelops melacanthus, euschistus heros, and podisus nigrispinus
  (hemiptera: Pentatomidae)}, Neotropical Entomology 47 (2018), pp. 543--552.

\bibitem{R22}
 {R Core Team}, \emph{R: A Language and Environment for Statistical Computing},
  R Foundation for Statistical Computing, Vienna, Austria (2022).
  \urlprefix\url{https://www.R-project.org/}.

\bibitem{Shimbori2023}
M.E. Shimbori1, B.R. Querino, A.V. Costa, and Z.A. R., \emph{Taxonomy and
  biological control: New challenges in an old relationship}, Neotropical
  Entomology  (2023), p.~22.

\bibitem{Therneau}
T.M. Therneau and P.M. Grambsch, \emph{Modeling Survival Data: Extending the
  Cox Model}, Springer, New York, 2000.

\bibitem{Tutz2012}
G. Tutz, \emph{Regression for categorical data}, Cambridge University Press,
  2012.

\bibitem{Ware1988}
S.F. Ware~JH Lipsitz~S, \emph{Issues in the analysis of repeated categorical
  outcomes}, Stat Med. 7-2 (1988), pp. 95--107.

\bibitem{wiedemann2003}
L.M. Wiedemann, C. Canto-Silva, H.P. Romanowski, and L.R. Redaelli,
  \emph{Oviposition behaviour of gryon gallardoi (hym.; scelionidae) on eggs of
  spartocera dentiventris (hem.; coreidae)}, Brazilian Journal of Biology 63
  (2003), pp. 133--139.

\bibitem{yao2022}
T. Yao, W. Qin, L. Meng, and B. Li, \emph{Oviposition and developmental
  performances of the gregarious hyperparasitoid cheiloneurus nankingensis in
  relation to host age}, Biological Control 172 (2022), p. 104967.

\bibitem{Zeger1992}
S.L. Zeger and K.Y. Liang, \emph{An overview of methods for the analysis of
  longitudinal data.}, Statistics in medicine 11 14-15 (1992), pp. 1825--39.
  \urlprefix\url{https://api.semanticscholar.org/CorpusID:38838230}.

\end{thebibliography}

\end{document}